# Shape measurement of single gold nanorods in water using open-access optical microcavities

Yumeng Yin, Aurelien Trichet, Jiangrui Qian, Jason Smith[*]

*Department of Materials, University of Oxford, 16 Parks Road, OX1 3PH, United Kingdom*

*\* jason.smith@materials.ox.ac.uk*

## Abstract

Shape measurement of rod-shaped particles in fluids is an outstanding challenge with applications in characterising synthetic functional nanoparticles and in early warning detection of rod-shaped pathogens in water supplies. However, it is challenging to achieve accurate and real-time measurements at a single particle scale in solution with existing methods. Here we introduce a novel technique to measure the aspect ratio of rod-shaped particles by analysing changes in the polarisation state of a laser beam transmitted through an optical microcavity through which the particle diffuses. The resolution in aspect ratio measurement is found to be around 1%. Our work opens the new possibility of in-situ and single-particle shape measurements, which have promising applications in nanoparticle characterisation, water monitoring, and beyond.

## 1  Introduction

Nanoparticles have in the past 20 years become vital tools in various fields, such as drug delivery[1], [2], [3] and catalysis[4], [5], [6]. The shape of a nanoparticle can significantly influence its properties and performance across these application domains. For example, in nanomedicine, rod-shaped particles allow for a higher drug loading quantity owing to their larger surface area to volume ratio[7] and they can also be advantageous for targeting specific tissues or sites[8]. The characterisation of particle shape is, therefore, an important capability for research and development, and for quality control in manufacturing.

Many natural pathogens, such as Legionella and Escherichia coli (E.coli) bacteria, are also rod-shaped, and the ability to identify their presence in water supplies is important for public health. Shape measurement can therefore provide an early indication of these harmful pathogens in fluids and helps avoid the infection and spreading of diseases.

Various approaches have been taken to measuring nanoparticle shape. If the particles can be isolated and a suitable sample preparation is carried out, electron microscopy (EM) provides detailed shape information via images with resolution at the nanometre-scale or higher. Fluid-based measurements of particle shape are more challenging due to rapid Brownian diffusion, which introduces a random walk to both the location and the orientation of the particles, typically on sub-millisecond time scales. For larger particles in the micrometre size regime, machine learning algorithms have been used for the shape-based classification of microparticles with moment-based holography[9]. However, the pixel size limits the size of detected particles, and shape features may not remain for regular shapes such as rods with the perspective and camera speed problem when dealing with smaller particles.

Particles made of different materials have properties that are shape-dependent, which, in turn, we can use to extract shape information. For example, metal nanorods display a plasmon resonance, the frequency of which depends on the aspect ratio and can be measured by UV-Vis optical absorption spectroscopy[10]. Dielectric particles, on the other hand, display no such resonance but have an anisotropic dielectric polarisability that depends weakly on shape[11], [12], [13]. Optical scattering techniques such as dynamic light scattering (DLS) and nanoparticle tracking analysis (NTA) have commercial products on the market and are convenient to use. Advanced depolarized DLS[14], provides shape information but requires particle concentrations of $10^{12}$ nanoparticles/ml for a typical commercial instrument and practically can only obtain averaged values for all the particles in fluids. NTA measures single particles but as an imaging technique, its sampling rate is generally too slow to capture rotational motion, and thus so far it has not been successful in providing quantitative shape measurement.

The diffusion of anisotropic particles has also been explored. From simple direct digital video microscopy imaging[15] to the analysis of scattered light with polarisation information[16], [17], constrained two-dimensional movement has been studied. It shows dependence on the local environment and again requires a high-speed advanced data capture method for faster movement. The three-dimensional, more complicated movement has also been investigated in a viscous environment[18] limited by the temporal resolution of the camera[19]. For example, single particle orientation and rotation tracking (SPORT) combines traditional single particle tracking (SPT) and Nomarski-type differential interference contrast (DIC)[20]. It can perform a systematic study of nanoscale rotational dynamics[21]. With the assistance of a convolutional neural network, it can track anisotropic gold nanoparticle probes in living cells[20]. However, to capture quicker translational and rotational movement of nanorods, more advanced methods must be put forward.

Optical microcavities provide a means to create a small, highly resonant optical mode sensitive to small perturbations and have been used previously to measure the polarisability and hydrodynamic diameter of individual spherical nanoparticles[22]. In this work, we show that a similar fluid-based measurement can be performed on single nanoparticles to determine their shape[23]. Aspect ratios $\mu$ of single gold nanorods are determined by measuring induced changes to the polarisation state of a probe laser beam transmitted through the microcavity. Analysing the Stokes parameters of the transmitted beam allows the extraction of polarisability-related anisotropy parameters and the derivation of $\mu$. The results obtained using this method are compared with those from scanning electron microscopy (SEM).

## 2  Measurement principles

The microcavities consist of opposing concave and planar mirrors with low-loss distributed Bragg reflector (DBR) coatings (see SI section 1) to form highly resonant modes (Q-factor~10,000)[24]. In this work, microcavities with concave mirror radii of curvature (RoC) of 25 µm and a cavity length of about 1.5 µm are used (Fig. 1(a)). Resonance with the incident laser beam ($\lambda = 640$ nm) is achieved when the cavity length satisfies $L = \frac{q\lambda}{2n_m}$ ($q$ is the mode number, $\lambda$ is the laser wavelength and $n_m$ is the refractive index of the medium).

The incident laser beam is circularly polarised to excite all linearly polarised modes equally. When a particle enters the microcavity, its interaction with the incident light changes the state of the incident light, leaving specific 'fingerprints' for us to decrypt. Therefore, we built a polarimeter shown in Fig. 1(c) to record three Stokes parameters of the transmitted light from the microcavity[25]: $S_0$, the intensity of the transmitted light collected by an avalanche photodiode (APD); $S_1$ and $S_3$, the degree of linear and circular polarisation of the transmitted light, both collected by a balanced photodiode separately (see SI section 1, 2).

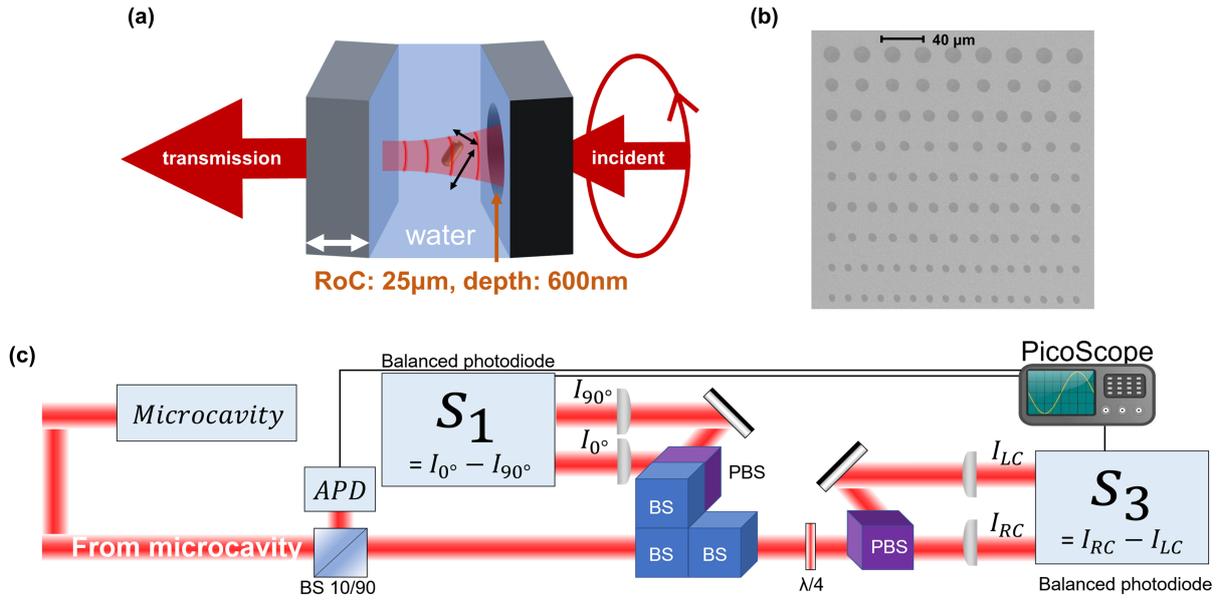

Figure 1. Schematic of the experimental setup. (a) Schematic of a gold nanorod diffusing in the microcavity. Double-headed black arrows indicate directions parallel and transverse to the rod axis. The double-headed white arrow shows the scanning of the planar mirror. (b) SEM image of an array of micromirrors of different sizes used for cavity construction. (c) Schematic of the polarimeter to analyse the transmitted light.

When a particle enters the microcavity, the polarisability of the particle relative to the water medium causes a spectral shift in the resonance. The cavity length is modulated at a frequency of 5 kHz, thus sweeping the cavity mode repeatedly through resonance. The transmitted light is recorded on the time scale and then converted to a function of frequency detuning $\Delta$ between the resonance with just the probe laser light and in the presence of a particle (see SI section 3). Since the polarisability of the nanorod is highly anisotropic, it splits the cavity resonance mode into two eigenstates parallel to the principal axes of the rod projected onto the cavity plane (X and Y in Fig. 2(a)). These two orthogonal linearly polarised eigenstates induce different phase shifts at the laser wavelength caused by differences in polarisabilities along these two directions. Thus, the transmitted light obtains a degree of linear polarisation which is a function of detuning.

To obtain quantitative information about this impact and thus derive the aspect ratio of the gold nanorods, we developed a theoretical model (see SI section 4). The Stokes parameters $S_0$ and $S_1$ can be expressed as functions of the detuning, $\Delta$, between the resonant frequency of the microcavity with just the probe laser light and in the presence of a spherical particle:

$$S_0(\Delta) = \frac{F^2}{(\Delta + \varepsilon)^2 + (\gamma + \eta)^2} + \frac{F^2}{(\Delta - \varepsilon)^2 + (\gamma - \eta)^2} \qquad (1a)$$

$$S_1(\Delta) = 4F^2 \frac{\gamma(\varepsilon \sin 2\psi - \eta \cos 2\psi) - \Delta(\eta \sin 2\psi + \varepsilon \cos 2\psi)}{(\Delta^2 - \varepsilon^2 + \gamma^2 - \eta^2)^2 + 4(\Delta\eta - \varepsilon\gamma)^2} \qquad (1b)$$

where $F$ is the amplitude of the incident field and $\gamma$ is the total photon loss rate from the cavity in the presence of a spherical particle. $2\varepsilon$ and $2\eta$ represent the particle anisotropy and are the differences in the resonant frequency and in the loss between the two orthogonal eigenstates. $\psi$ is the azimuthal angle of the nanorod about the optical axis of the cavity as referred to the instrument axes (Fig. 2(a)).

Here $S_1$ provides the most sensitive measurement of $\varepsilon$ and $\eta$ since its amplitude is proportional to these parameters, but data analysis is complicated by the dependence on $\psi$. Alternatively, with sufficient anisotropy that the mode splitting is comparable in magnitude to the mode width, $\varepsilon$ and $\eta$ can be established from $S_0$ alone.

With the anisotropy-related parameters $\varepsilon$ and $\eta$ determined, it is convenient to define an 'anisotropy loss tangent' $\Phi = \frac{\eta}{\varepsilon}$ which can be evaluated for each measurement of $S_0(\Delta)$ to derive the particle aspect ratio. For small particles, in which losses are dominated by absorption and scattering can be neglected, it is independent of both the position and the orientation of the nanorod within the cavity mode. It is only determined by the particle itself. For larger particles, scattering is not negligible and $\Phi$ depends on both orientation and position of the particle in the microcavity (see SI section 6), but shape information can still be deduced with certain assumptions. In either case, the aspect ratio $\mu$ is readily deduced from $\Phi$ (see SI section 5, 6).

In this work, we evaluated this technique using two different types of gold nanorod samples A and B with different nominal aspect ratios (see SI section 1). The theoretical absorption and scattering cross sections for samples A and B[11], [12], [13] were calculated, using particle volumes estimated from SEM, to estimate the degree to which scattering can be neglected for these samples. Absorption was calculated to be 73% of the overall extinction for sample A and 82% for sample B.

## 3  Single particle detection events

When a particle enters the cavity, the increase in intra-cavity losses broadens the resonant mode peak such that the peak width provides a convenient signature of the particle's presence. Fig. 2(b) shows an example data set in which two discrete single-particle events are observed. Each event provides multiple measurements of $S_0(\Delta)$ and $S_1(\Delta)$, with both spectra changing as the particle diffuses in the cavity mode. That $S_1$ remains zero for spherical particles is verified by injecting gold nanospheres with a diameter of 60 nm into the apparatus (see SI section 7). When gold nanorod samples A and B are injected, $S_1(\Delta)$ shows clear features. In the data presented below however $\varepsilon$ and $\eta$ are determined from $S_0(\Delta)$. With these two parameters determined, $S_1$ is then used to obtain the azimuthal angle $\psi$.

Fig. 2(c)-(f) show examples of individual fits to $S_0(\Delta)$ and $S_1(\Delta)$ signals during the second single-particle event in Fig. 2(b). As the nanorod undergoes rotational diffusion, changes in polar angle $\theta$ translate to changes in the magnitude of the splitting of the $S_0$ peak and the amplitude of the $S_1$ response, while changes in $\psi$ translate to changes only in the shape of the $S_1$ response.

The width of the resonant peaks reveals that each measurement takes about 1 microsecond. The angular displacement caused by the free Brownian diffusion of the particles and optical torques are small within this time duration and so the particle can be considered stationary for the purposes of analysing each peak (see SI section 8). The temperature change on the gold nanorod surface as a result of optical absorption is also estimated, from which we conclude that particle shape change due to heat is negligible (see SI section 9).

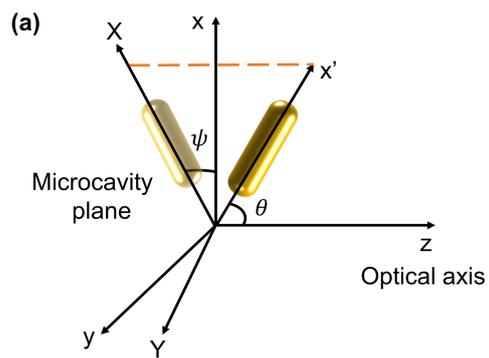
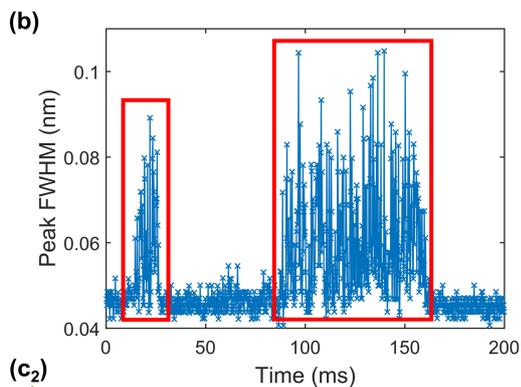
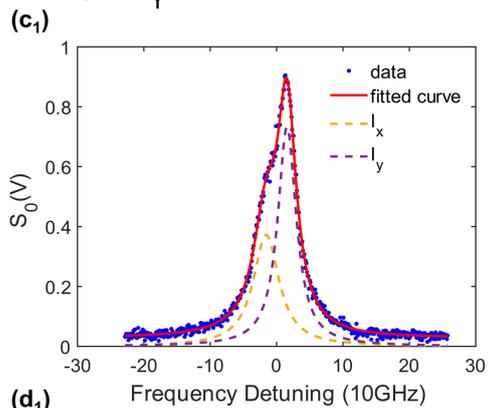
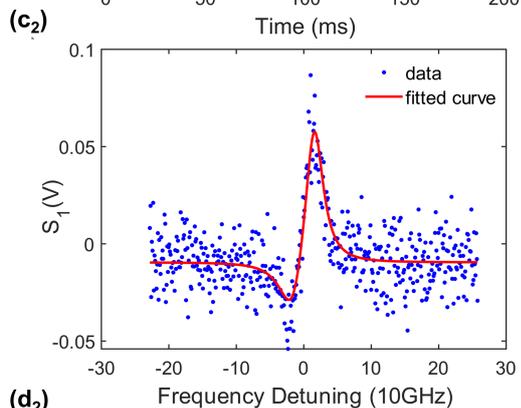
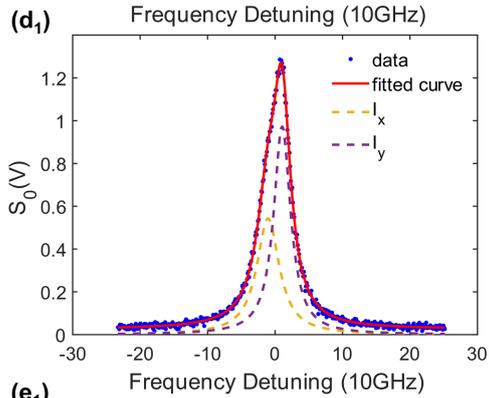
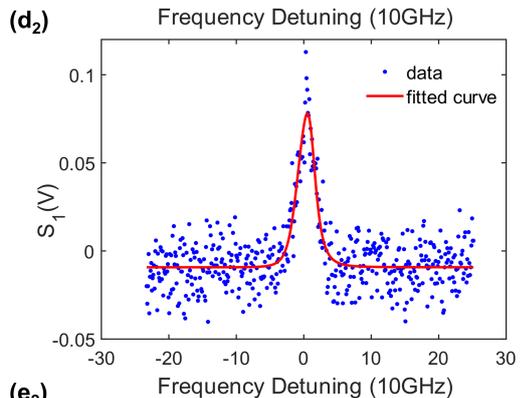
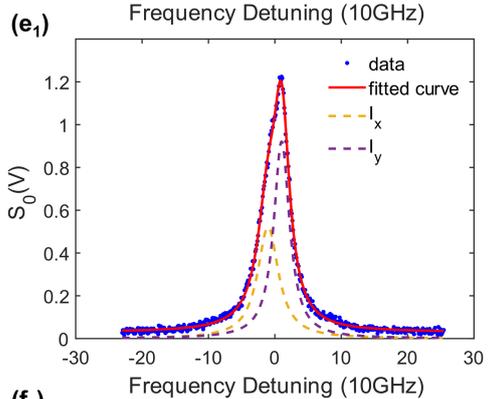
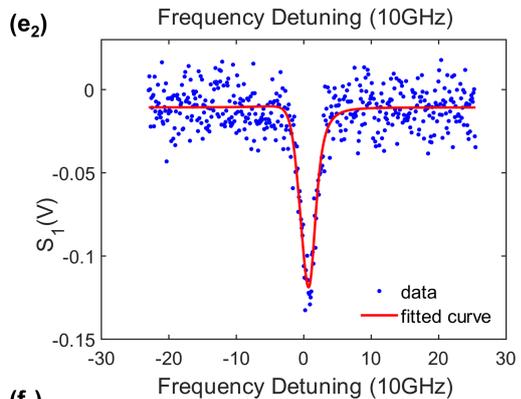
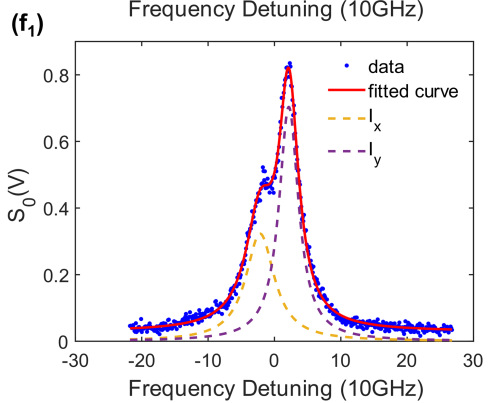
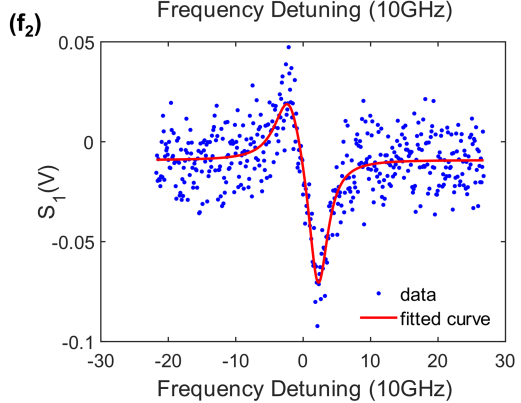

Figure 2. A single-particle event. (a) Schematic of the coordinate system used. (b) FWHM of the $S_0$ peak measured using the APD over a 200 ms window, showing two discrete single-particle events due to loss-induced broadening. (c)-(f) Examples of individual $S_0$ and $S_1$ peaks during the second event. Blue dots are raw data and red curves are fitted curves using equations (1a) and (1b).

The histograms for $\psi$ and $\Phi$ for a single-particle event from each of samples A and B are shown in Fig. 3(a)-(d), and Fig.3(e) illustrates the theoretical relationship between $\mu$ and $\Phi$ (see SI section 10). In each case, the $\psi$ distribution is flat indicating no preferred orientation of the particle in the lab frame. The $\Phi$ distributions in each case show a clear single peak in the range $0.2 < \Phi < 0.3$. Neglecting scattering and fitting a normal distribution results in $\mu = 2.00 \pm 0.02$ and $\mu = 1.87 \pm 0.02$ for the sample A and B particles respectively. The magnitudes of the standard errors in $\mu$ suggest a basic measurement resolution of around 1%. Including scattering and fitting using the expected $\Phi$ distribution yields smaller values, $\mu = 1.66 \pm 0.05$ and $\mu = 1.69 \pm 0.04$ for the sample A and B particles respectively (see SI section 6). The degree of scattering depends on particle volume which, in this work, is estimated from SEM data. It also depends on the optical density of states in the microcavity, which is lower than that of free space and is a function of position within the cavity mode. A detailed analysis of scattering is beyond the scope of this work – the red curves in Fig. 3(b) and Fig. 3(d) assume the same density of states as in free space.

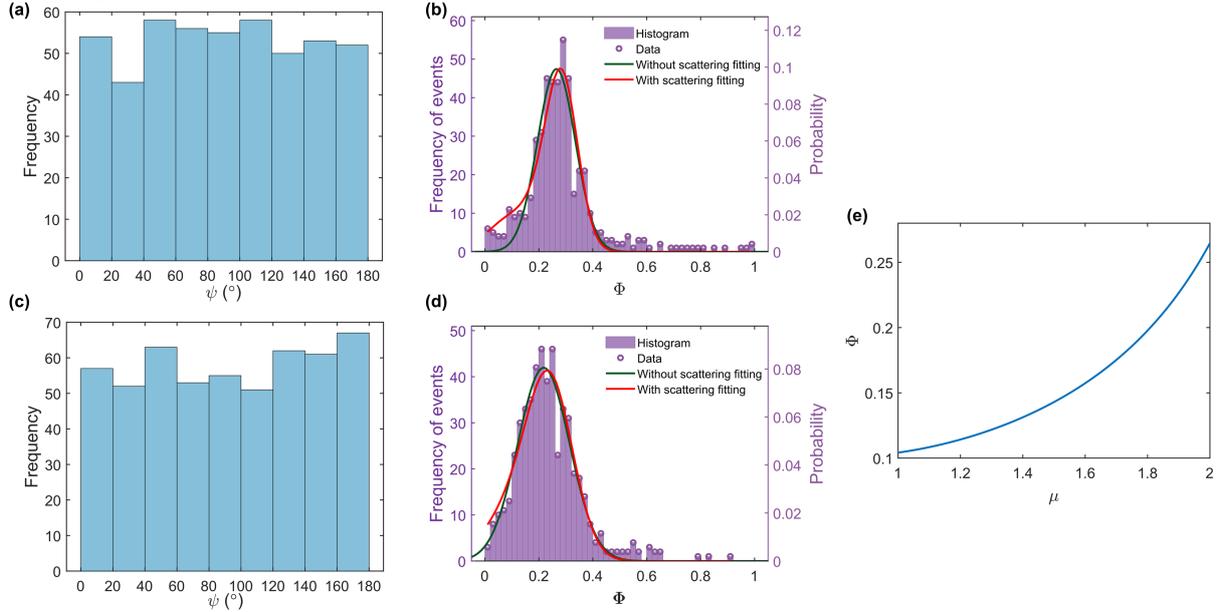

Figure 3. Summary of a single-particle event. Sample A: (a), (b); Sample B: (c), (d). (a), (c) Histograms of azimuthal angle $\psi$. (b), (d) Histograms of anisotropy loss tangent $\Phi$. The green curve is the fitting without scattering while the red curve is the fitting with scattering using SEM volume. (e) The numerical relationship between $\mu$ and $\Phi$.

## 4 Statistical results for 30 particles

Fig. 4(a) shows the measured aspect ratios, neglecting scattering effects, for 30 particles in each sample. The mean values and standard deviations are $\mu = 1.85 \pm 0.13$ and $\mu = 1.89 \pm 0.11$ for samples A and B respectively. SEM images of nanoparticles from the two samples were used to provide an independent measure of $\mu$, resulting in the histograms shown in Fig.4(b). Although the mean values produced by the two methods differ, the distributions are similar with samples A and B overlapping in both cases. If scattering is included in the analysis using the averaged volume calculated from the SEM images, the aspect ratios of the 30 particles in each sample become $\mu = 1.4 \pm 0.2$ and $\mu = 1.7 \pm 0.14$. The nanorods in sample A are larger than those in sample B, which induces a more pronounced effect of scattering in reducing the calculated values of $\mu$. When the scattering is included, the anisotropy loss tangent is related to the particle's position due to the difference in the optical density of states in the cavity compared with the free space[22]. If the scattering is reduced to 0.65 times that of the free space[22] the calculated aspect ratios become $1.59 \pm 0.16$ and $1.82 \pm 0.12$ respectively.

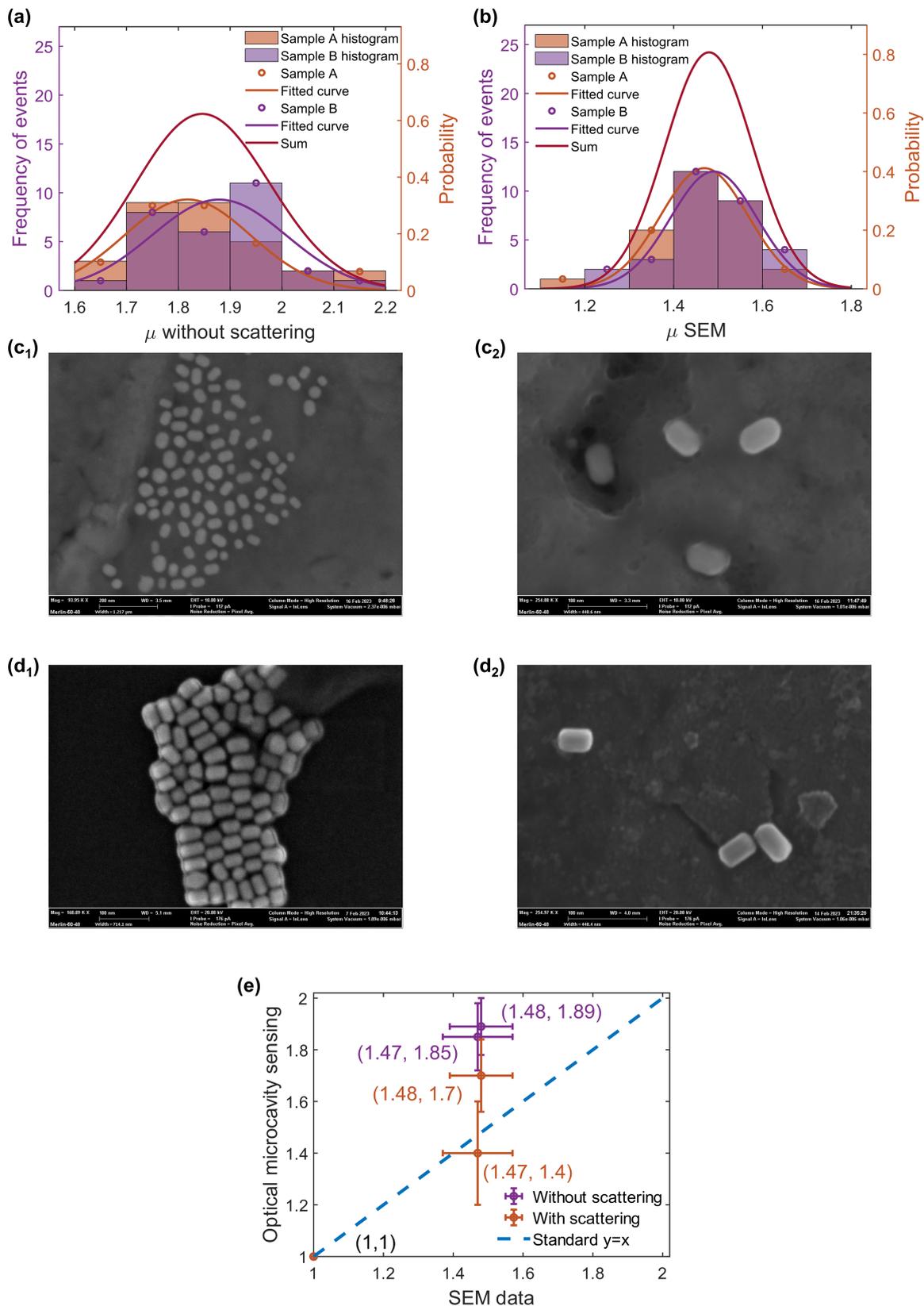

Figure 4. Summary of statistical results of 30 particles in sample A and B. (a) Optical microcavity sensing method results without scattering. (b) SEM results directly measured from SEM images. (c), (d) SEM images of samples A and B. $c_2$ and $d_2$ are images with higher magnification. (e) Relationship between optical microcavity sensing results and the SEM results.

# 5 Discussion

Table 1 compares the aspect ratios of 30 particles for the two samples each as determined by the different methods used. The aspect ratios calculated from SEM data depend on the assumed distribution of particle orientations – we present data assuming particles lying flat in the image plane and for particles assuming randomly oriented (see SI section 11). The two samples provide very similar results, with aspect ratios between 1.5 to 1.6. The values obtained using optical microcavity sensing neglecting scattering are somewhat larger than those obtained from the SEM measurements at around 1.9, but again with similar aspect ratios for samples A and B. When scattering is considered, the aspect ratios reduce to 1.4 and 1.7 respectively if one assumes no reduction in the optical density of states, and 1.6 to 1.8 if one assumes an average optical density of states reduced to 0.65 of the free space value. One thing to mention here is that the 30 particles for SEM and optical microcavity sensing are not exactly the same, because the diffusion in the flow cell is random, and we could not directly map it to any specific particle in SEM images. We would expect when the number of particles increases, the accuracy will also increase accordingly.

| Method | $\mu$ of sample A | $\mu$ of sample B |
|---|---|---|
| SEM assuming rods lay flat | $1.47 \pm 0.1$ | $1.48 \pm 0.09$ |
| SEM assuming random orientation | $1.56 \pm 0.17$ | $1.57 \pm 0.16$ |
| Optical microcavity sensing (without scattering) | $1.85 \pm 0.13$ | $1.89 \pm 0.11$ |
| Optical microcavity sensing (with scattering, SEM volume, $\beta(\vec{r}) = 1$) | $1.4 \pm 0.2$ | $1.7 \pm 0.14$ |
| Optical microcavity sensing (with scattering, SEM volume, $\beta(\vec{r}) = 0.65$) | $1.59 \pm 0.16$ | $1.82 \pm 0.12$ |

Table 1: Statistical comparison with other methods

Note that the volume factors used here are based on average volume: they can in principle be established on a particle-by-particle basis within the same microcavity analysis apparatus, using either the magnitude of the mode shift[22] or the temporal autocorrelation function. Indeed these two parameters can be combined to establish the refractive index of the particle at the probe wavelength, removing the need for independent knowledge of the particle composition. While advantageous in providing a standalone single particle shape measurement, this approach would be unlikely to change the general form of the results presented here, and the reason for the slight discrepancy between the SEM measurements and the optical measurements remains a subject of further investigation. We also performed a theoretical analysis on exploring the measuring limit of the minimum aspect ratio with our apparatus (see SI section 12). With the appropriate assumptions, the minimum number is around 1.05 in this case.

# 6 Conclusions

In summary, we propose measuring the aspect ratio of gold nanorods using optical microcavity sensing. By measuring Stokes parameters of the transmitted light through the microcavity with gold nanorods, the polarisability anisotropy can be deduced to obtain the shape anisotropy. The method presented here can be adapted to dielectric nanoparticles such as rod-shaped pathogens. For other complex shapes, not limited to regular shapes such as a rod, like a disk or triangle shape, this technique could still be adapted accordingly to perform more advanced analysis.

## Acknowledgements
The authors are grateful to Zhin Mai and Gareth Jones for their discussions and Zhin Mai for comments on the manuscript.

## Research Funding Statement
The authors acknowledge the use of characterisation facilities within the David Cockayne Centre for Electron Microscopy, Department of Materials, University of Oxford, alongside financial support provided by the Henry Royce Institute (Grant ref EP/R010145/1).

## Author Contribution Statement
J.S. and A.T. conceived the idea and initiated the project. Y.Y built the setup, performed the experiments, analysed the data and wrote the paper under the supervision of J.S. J.Q. fabricated the microcavity arrays. All authors have accepted responsibility for the entire content of this manuscript and approved its submission.

## Conflict of Interest Statement
Authors state no conflicts of interest.

## Data Availability
The datasets generated and/or analysed during the current study are available from the corresponding author upon reasonable request.

## References

[1] M. J. Mitchell, M. M. Billingsley, R. M. Haley, M. E. Wechsler, N. A. Peppas, and R. Langer, "Engineering precision nanoparticles for drug delivery," *Nat Rev Drug Discov*, vol. 20, no. 2, pp. 101–124, 2021, https://doi.org/10.1038/s41573-020-0090-8.

[2] E. Blanco, H. Shen, and M. Ferrari, "Principles of nanoparticle design for overcoming biological barriers to drug delivery," *Nat Biotechnol*, vol. 33, no. 9, pp. 941–951, 2015, https://doi.org/10.1038/nbt.3330.

[3] de Jong, "Drug delivery and nanoparticles: Applications and hazards," *Int J Nanomedicine*, vol. 3, no. 2, pp. 133–149, 2008, https://doi.org/10.2147/IJN.S596.

[4] T. Li *et al.*, "Denary oxide nanoparticles as highly stable catalysts for methane combustion," *Nat Catal*, vol. 4, no. 1, pp. 62–70, 2021, https://doi.org/10.1038/s41929-020-00554-1.

[5] C. Xie, Z. Niu, D. Kim, M. Li, and P. Yang, "Surface and Interface Control in Nanoparticle Catalysis," *Chem Rev*, vol. 120, no. 2, pp. 1184–1249, 2020, https://doi.org/10.1021/acs.chemrev.9b00220.

[6] D. Astruc, "Introduction: Nanoparticles in Catalysis," *Chem Rev*, vol. 120, no. 2, pp. 461–463, 2020, https://doi.org/10.1021/acs.chemrev.8b00696.

[7]  Y. You, L. He, B. Ma, and T. Chen, "High-Drug-Loading Mesoporous Silica Nanorods with Reduced Toxicity for Precise Cancer Therapy against Nasopharyngeal Carcinoma," *Adv Funct Mater*, vol. 27, no. 42, p. 1703313, 2017, https://doi.org/10.1002/adfm.201703313.

[8]  V. P. Chauhan *et al.*, "Fluorescent nanorods and nanospheres for real-time in vivo probing of nanoparticle shape-dependent tumor penetration," *Angewandte Chemie - International Edition*, vol. 50, no. 48, pp. 11417–11420, 2011, https://doi.org/10.1002/anie.201104449.

[9]  H. Luo, J. Xu, J. Jiao, L. Zhong, X. Lu, and J. Tian, "Moment-Based Shape-Learning Holography for Fast Classification of Microparticles," *Adv Photonics Res*, vol. 4, no. 8, 2023, https://doi.org/10.1002/adpr.202300120.

[10] R. Kumar *et al.*, "Determination of the Aspect-ratio Distribution of Gold Nanorods in a Colloidal Solution using UV-visible absorption spectroscopy," *Sci Rep*, vol. 9, no. 1, p. 17469, 2019, https://doi.org/10.1038/s41598-019-53621-4.

[11] R. Gans, "Über die Form ultramikroskopischer Goldteilchen," *Ann Phys*, vol. 342, no. 5, pp. 881–900, 1912, https://doi.org/10.1002/andp.19123420503.

[12] R. Gans, "Über die Form ultramikroskopischer Silberteilchen," *Ann Phys*, vol. 352, no. 10, pp. 270–284, 1915, https://doi.org/10.1002/andp.19153521006.

[13] G. C. Papavassiliou, "Optical properties of small inorganic and organic metal particles," *Progress in Solid State Chemistry*, vol. 12, no. 3–4, pp. 185–271, 1979, https://doi.org/10.1016/0079-6786(79)90001-3.

[14] M. Glidden and M. Muschol, "Characterizing gold nanorods in solution using depolarized dynamic light scattering," *Journal of Physical Chemistry C*, vol. 116, no. 14, pp. 8128–8137, 2012, https://doi.org/10.1021/jp211533d.

[15] Y. Han, A. M. Alsayed, M. Nobili, J. Zhang, T. C. Lubensky, and A. G. Yodh, "Brownian Motion of an Ellipsoid," *Science*, vol. 314, no. 5799, pp. 626–630, 2006, https://doi.org/10.1126/science.1130146.

[16] M. Mazaheri, J. Ehrig, A. Shkarin, V. Zaburdaev, and V. Sandoghdar, "Ultrahigh-Speed Imaging of Rotational Diffusion on a Lipid Bilayer," *Nano Lett*, vol. 20, no. 10, pp. 7213–7219, 2020, https://doi.org/10.1021/acs.nanolett.0c02516.

[17] F. Hajizadeh, L. Shao, D. Andrén, P. Johansson, H. Rubinsztein-Dunlop, and M. Käll, "Brownian fluctuations of an optically rotated nanorod," *Optica*, vol. 4, no. 7, pp. 746–751, 2017, https://doi.org/10.1364/optica.4.000746.

[18] K. Chaudhari and T. Pradeep, "Spatiotemporal mapping of three dimensional rotational dynamics of single ultrasmall gold nanorods," *Sci Rep*, vol. 4, no. 1, pp. 5948–5948, 2014, https://doi.org/10.1038/srep05948.

[19] T. Fordey *et al.*, "Single-Shot Three-Dimensional Orientation Imaging of Nanorods Using Spin to Orbital Angular Momentum Conversion," *Nano Lett*, vol. 21, no. 17, pp. 7244–7251, 2021, https://doi.org/10.1021/acs.nanolett.1c02278.

[20] D. Song *et al.*, "Deep Learning-Assisted Automated Multidimensional Single Particle Tracking in Living Cells," *Nano Lett*, vol. 24, no. 10, pp. 3082–3088, 2024, https://doi.org/10.1021/acs.nanolett.3c04870.

[21] Y. Sun *et al.*, "Correlation and Autocorrelation Analysis of Nanoscale Rotational Dynamics," *The Journal of Physical Chemistry C*, vol. 127, no. 15, pp. 7327–7334, 2023, https://doi.org/10.1021/acs.jpcc.3c00286.

[22] A. A. P. Trichet, P. R. Dolan, D. James, G. M. Hughes, C. Vallance, and J. M. Smith, "Nanoparticle Trapping and Characterization Using Open Microcavities," *Nano Lett*, vol. 16, no. 10, pp. 6172–6177, 2016, https://doi.org/10.1021/acs.nanolett.6b02433.

[23] Y. Yin, A. Trichet, and J. Smith, "Aspect ratio measurement of water-based rod-shaped particles in optical microcavities," SPIE, 2024, https://doi.org/10.1117/12.3001900.

[24] P. R. Dolan, G. M. Hughes, F. Grazioso, B. R. Patton, and J. M. Smith, "Femtoliter tunable optical cavity arrays," *Opt Lett*, vol. 35, no. 21, pp. 3556–3558, 2010, https://doi.org/10.1364/ol.35.003556.

[25] S. Shibata, N. Hagen, S. Kawabata, and Y. Otani, "Compact and high-speed Stokes polarimeter using three-way polarization-preserving beam splitters," *Appl Opt*, vol. 58, no. 21, pp. 5644–5649, 2019, https://doi.org/10.1364/ao.58.005644.

# Shape measurement of single gold nanorods in water using open-access optical microcavities

# Supporting Information

Yumeng Yin, Aurelien Trichet, Jiangrui Qian, Jason Smith[*]

*Department of Materials, University of Oxford, 16 Parks Road, OX1 3PH, United Kingdom*

Table of contents

## 1. Experimental setup

A complete schematic of the optical apparatus is shown in Figure S1. The intensity of the laser light is adjustable with the combination of a half-wave plate and a polarising beam splitter (PBS). Three parts of the apparatus are identified in blue boxes. Part 1 splits the laser beam into two beams using a half-waveplate and two PBSs. The intensities and alignment of the two beams can be adjusted independently. One beam is the sensing beam to perform the particle characterisation, while the other one is the reference beam to remove background noise. Part 2 of the apparatus includes the microcavity assembly and an imaging system used both for alignment and to perform cavity locking through a proportional-integral-derivative (PID) control system. It provides a locking function by monitoring fringes formed by the transmission of off-resonance LED light through the planar side of the cavity. Part 3 is the polarimeter used to analyse the polarisation state of the transmitted laser light. Within the polarimeter, two 50:50 beam splitters (BS) remove the linear retardance and diattenuation to preserve the original polarisation state of the light. Polarising beam splitters (PBS) split the light into orthogonal components and balanced photodiodes are used to measure intensity differences between them. Picoscope gets triggered by signs of single-particle events and collects the voltage data from the avalanche photodiode (APD) and two balanced photodiodes as a function of time simultaneously for further data analysis.

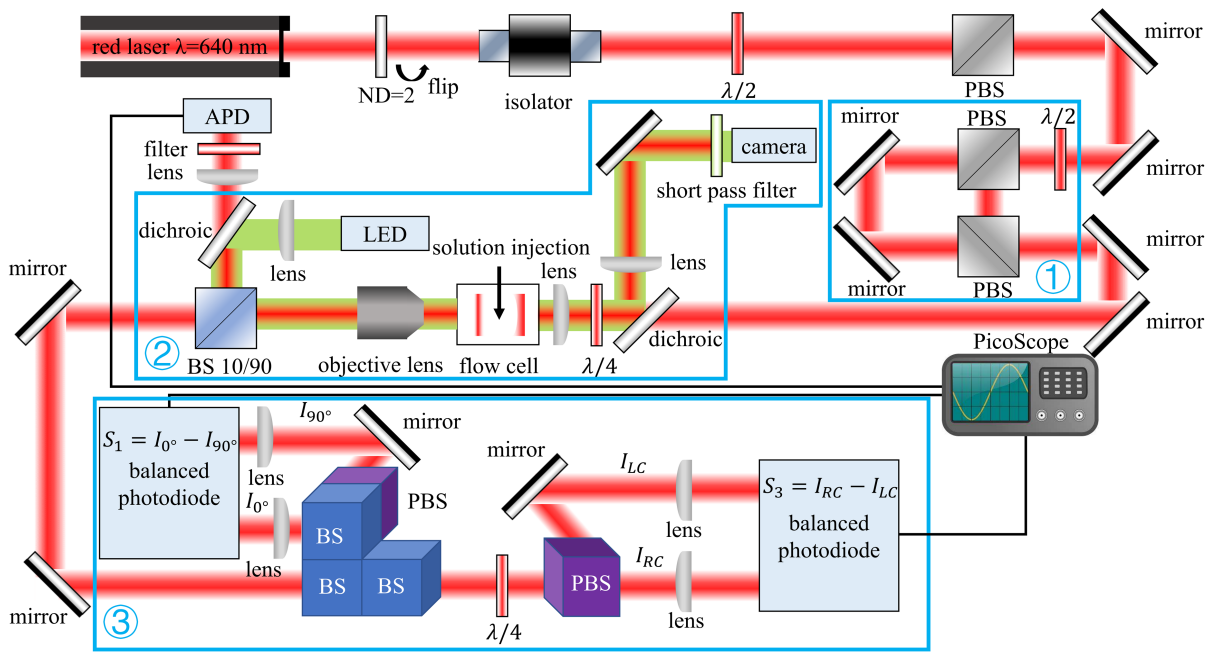

Figure S1: Schematic of the experimental setup. Part 1 prepares two beams for the experiment. Part 2 includes the flow cell and the imaging system for the microcavity array. Part 3 is the polarimeter measuring Stokes parameters.

A microfluidic system was used to inject the solution into the microcavity. Gold nanorods in aqueous solution (Nanopartz Inc.) of two designs (samples A and B) were tested. Diluted concentrations of around $10^9$ nanoparticles /mL provided single particle events every few seconds. Sample volumes of 200-300 µL were used for each measurement, determined by the dimensions of the flow cell and microfluidic assembly. Upon sample injection, as soon as the first particle was observed entering the cavity, the flow was stopped and the particles in the solution were allowed to diffuse freely within the flow cell reservoir.

Substrates for concave mirrors were prepared using focused ion beam milling. Mirrors were $SiO_2/Ta_2O_5$ Bragg stacks with reflectivity of 99.99% and 99.92% for concave and planar sides (Layertec GmbH).

## 2. Polarimeter validation

Before inserting the flow cell with the cavity, the polarimeter was used to measure Stokes parameters of a prepared laser beam to check its performance. The quarter-waveplate in part 2 of the apparatus (Figure S1) was used to prepare the incident light with different polarisation states by rotating at different angles. The Stokes parameters were normalised:

$$\widetilde{S_1} = \frac{I_{0°} - I_{90°}}{I_{0°} + I_{90°}} \quad (1a)$$

$$\widetilde{S_3} = \frac{I_{RC} - I_{LC}}{I_{RC} + I_{LC}} \quad (1b)$$

These measured values were then compared with the standard theoretical values[1] and the difference was plotted as a function of the rotating angle. Measurements were performed both directly using a power meter and after fibre coupling. The results in Figure S2 show that the measured polarisation state agrees with theoretical prediction to within 5% across all measurements.

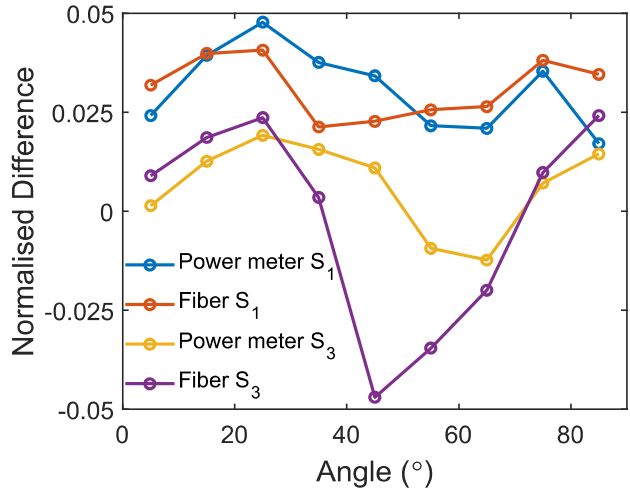

Figure S2: Polarimeter validation. Normalised Stokes parameters $S_1$ and $S_3$ were measured with different rotating angles of the quarter-waveplate. Differences with theoretical values were plotted.

## 3. Conversion between time and frequency domain

With the scanning at 5 kHz, data is recorded in the time domain. It is then converted to the frequency domain:

$$\omega = \frac{2\pi c}{\lambda} \quad (2a)$$

whereby differentiating gives:

$$\frac{\Delta}{\omega_c} = -\frac{\Delta \lambda}{\lambda} = -\frac{2n_m}{q}\frac{\Delta L}{\lambda} \quad (2b)$$

where $\Delta$ is the frequency detuning, $\omega_c$ is the frequency for the probe laser light, $n_m$ is the refractive index of the media, $q$ is the mode number, and $\lambda = 640\ nm$ is the wavelength of the probe laser. $\Delta L$ is achieved by the sinusoidal wave scanning with the piezoelectric stacks glued on the planar mirror side. Therefore, the frequency detuning is:

$$\Delta= -\frac{n_m w_c}{q\lambda} * scan\ range * [\sin(0.01\pi(t-t_0)) - \sin(0.01\pi(t_{mid}-t_0))] \qquad (3)$$

where $t_{mid}$ corresponds to the middle point of the two modes, $t_0$ is the delay of 0 scanning point in reference to the data set collection start point. The scan range is determined by the piezoelectric response of the stack.

### 4. Stokes parameters derivation

The incident light was prepared with a left-hand circular polarisation. The electric fields in the transmitted beam, along the directions parallel (X) and perpendicular (Y) to the major axis of the projected rod (main text Fig. 2(a)), can then be written as:

$$e_X = \frac{Fe^{-i\omega t}}{(\Delta+\varepsilon)+i(\gamma+\eta)} \quad ; \quad e_Y = \frac{iFe^{-i\omega t}}{(\Delta-\varepsilon)+i(\gamma-\eta)} \qquad (4)$$

where $F$ is the amplitude of the incident field, $\Delta$ represents the detuning between the resonant frequency of the cavity with just the laser and in the presence of a spherical particle, and $\gamma$ is the total photon loss rate in the cavity in the presence of a spherical particle. $2\varepsilon$ and $2\eta$ represent the particle anisotropy and are the differences in the resonant frequency and in the loss between the two orthogonal eigenstates. Four Stokes parameters are shown below:

$$S_{00} = |e_X|^2 + |e_Y|^2 \qquad (5a)$$
$$S_{10} = |e_X|^2 - |e_Y|^2 \qquad (5b)$$
$$S_{20} = \frac{1}{2}|e_X+e_Y|^2 - \frac{1}{2}|e_X-e_Y|^2 \qquad (5c)$$
$$S_{30} = \frac{1}{2}|e_X+ie_Y|^2 - \frac{1}{2}|e_X-ie_Y|^2 \qquad (5d)$$

For a nanorod oriented at an azimuthal angle $\psi$ between the projected nanorod frame of reference (X, Y) and the instrument frame of reference (x, y) (main text Fig. 2(a)), the Stokes parameters in the instrument frame are then calculated by

$$\begin{pmatrix} S_0 \\ S_1 \\ S_2 \\ S_3 \end{pmatrix} = \begin{pmatrix} 1 & 0 & 0 & 0 \\ 0 & -\cos2\psi & \sin2\psi & 0 \\ 0 & -\sin2\psi & \cos2\psi & 0 \\ 0 & 0 & 0 & 1 \end{pmatrix} \begin{pmatrix} S_{00} \\ S_{10} \\ S_{20} \\ S_{30} \end{pmatrix} \qquad (6a)$$

$$S_0(\Delta) = \frac{F^2}{(\Delta+\varepsilon)^2+(\gamma+\eta)^2} + \frac{F^2}{(\Delta-\varepsilon)^2+(\gamma-\eta)^2} \qquad (6b)$$

$$S_1(\Delta) = 4F^2 \frac{\gamma(\varepsilon\sin2\psi - \eta\cos2\psi) - \Delta(\eta\sin2\psi + \varepsilon\cos2\psi)}{(\Delta^2-\varepsilon^2+\gamma^2-\eta^2)^2 + 4(\Delta\eta-\varepsilon\gamma)^2} \qquad (6c)$$

$$S_3(\Delta) = -2F^2 \frac{\Delta^2-\varepsilon^2+\gamma^2-\eta^2}{(\Delta^2-\varepsilon^2+\gamma^2-\eta^2)^2 + 4(\Delta\eta-\varepsilon\gamma)^2} \qquad (6d)$$

The measured $S_0(\Delta)$ and $S_1(\Delta)$ traces are analysed to provide the desired aspect ratio information. Since $S_3(\Delta)$ provides the same information as $S_0(\Delta)$, we didn't perform curve fitting for $S_3(\Delta)$. For the experiment, since the APD and balanced photodiodes (Figure S1) have different responsivities for light intensity, their voltage amplitude varies. However, this can be incorporated into different effective $F$ in the curve fitting process, and won't have any impact on our final results.

### 5. Polarisability anisotropy

The polarisability of a gold nanorod is described as a complex second-rank tensor. In the nanorod's frame of reference, the tensor takes a diagonal form:

$$\alpha = \begin{pmatrix} \alpha_W & 0 & 0 \\ 0 & \alpha_W & 0 \\ 0 & 0 & \alpha_L \end{pmatrix} \qquad (7)$$

where $\alpha_W$ and $\alpha_L$ are the polarisabilities along the short and long axis of the nanorod. Gans modification to Mie theory[2], [3], [4] relates the polarisabilities to the particle aspect ratio $\mu$ as follows:

$$\alpha_{W,L} = \frac{\epsilon - \epsilon_m}{\epsilon_m + p_{W,L}(\epsilon - \epsilon_m)} \tag{8}$$

$$p_L = \frac{1}{\mu^2 - 1}\left(\frac{\mu}{2\sqrt{\mu^2-1}}\ln\left(\frac{\mu+\sqrt{\mu^2-1}}{\mu-\sqrt{\mu^2-1}}\right) - 1\right), \qquad p_W = \frac{1-p_L}{2} \tag{9}$$

where $\epsilon$ is the permittivity of gold[5] at probe laser wavelength $\lambda = 640\ nm$, $\epsilon_m$ is the permittivity of the surrounding medium, and $p_{W,L}$ are the depolarisation factors. The complex nature of $\epsilon$ results in $\alpha_{W,L}$ being complex also. Projecting the polarisability tensor onto the plane transverse to the optical axis of the apparatus gives

$$\alpha' = \begin{pmatrix} \alpha'_L & 0 \\ 0 & \alpha'_W \end{pmatrix} = \begin{pmatrix} \cos^2\theta\ \alpha_W + \sin^2\theta\ \alpha_L & 0 \\ 0 & \alpha_W \end{pmatrix} \tag{10}$$

where $\theta$ is the angle of the nanorod to the optical axis (main text Fig. 2(a)). $\alpha'_L$ and $\alpha'_W$ is the polarisability along X and Y axis (main text Fig. 2(a)). The real and imaginary parts of $\alpha'$ result in shifts in the mode position and increases in mode width respectively.

## 6. Anisotropy loss tangent derivation and probability distribution

The anisotropy loss tangent is defined as $\Phi = \frac{\eta}{\varepsilon}$ in the main text:

$$\Phi = \frac{\eta}{\varepsilon} = \frac{\frac{w_c(\Delta\alpha_i + \Delta scattering\ part)}{4V_m}I(\vec{r})}{\frac{w_c\Delta\alpha_r}{4V_m}I(\vec{r})} \tag{11a}$$

where $w_c$ is the light frequency at the probe laser wavelength $640\ nm$, $V_m$ is the mode volume, $I(\vec{r})$ is the relative intensity at the particle position normalised to the strongest intensity in the cavity, $\Delta\alpha_i$, $\Delta\alpha_r$, and $\Delta scattering\ part$ are differences in the imaginary and real part of the polarisability and in the scattering part along the projected long and short axis of the gold nanorod (see X and Y direction in main text Fig. 2(a)). For small particles, in which losses are dominated by absorption, and scattering can be neglected, the anisotropy loss tangent $\Phi$ is:

$$\Phi = \frac{Im(\alpha'_L - \alpha'_W)}{Re(\alpha'_L - \alpha'_W)} = \frac{Im(\alpha_L - \alpha_W)}{Re(\alpha_L - \alpha_W)} \tag{11b}$$

It turns out to be a property of the particle. It is independent of both the position and orientation of the particle in the cavity mode.

For larger particles, scattering is not negligible. The photon loss rate introduced by scattering from the particle is:

$$\Gamma = \frac{\sigma_{sca}*I(\vec{r})}{A_{mode}}\frac{2}{\tau_{round\ trip}} = \frac{\sigma_{sca}}{A_{mode}}\frac{2I(\vec{r})}{\frac{2L}{c/n_m}} = \frac{\sigma_{sca}c}{n_m V_m}I(\vec{r}) = \frac{\omega_c}{2V_m}\frac{8\pi^2 n_m^3 V_p}{3\lambda^3}\alpha^2 I(\vec{r}) \tag{11c}$$

where $\sigma_{sca}$ is the scattering cross section of the particle, $A_{mode}$ is the mode cross section, $I(\vec{r})$, $V_m$ and $w_c$ are the same as in equation (11a), $\tau_{round\ trip}$ is the time of a round trip, $L$ is the cavity length, $n_m$ is the refractive index of the media, $c$ is the speed of light. It should be noted that I extracted $V_p$ from $\alpha^2$ and $\alpha^2 \propto V_p$ so that it has the same unit as the absorption factor. Therefore, the loss difference is $2\eta = \frac{\omega_c \Delta\alpha_i}{2V_m}I(\vec{r}) + \frac{\omega_c}{2V_m}\frac{8\pi^2 n_m^3 V_p}{3\lambda^3}\Delta(\alpha^2)I(\vec{r})$, and we define a volume-related factor $A = \frac{8\pi^2 n_m^3 V_p}{3\lambda^3}$.

The anisotropy loss tangent becomes:

$$\Phi = \frac{Im(\alpha_L-\alpha_W)+A\beta(\vec{r})\{\sin^2\theta\,\alpha_L^2-(\cos^2\theta+1)\alpha_W^2+2\cos^2\theta[Re(\alpha_L)Re(\alpha_W)+Im(\alpha_L)Im(\alpha_W)]\}}{Re(\alpha_L-\alpha_W)} \quad (11d)$$

where $\theta$ is the same as in equation (10). As we can see, now the anisotropy loss tangent is dependent both on the particle's position and on its orientation within the cavity mode. The positional dependence comes about because the optical density of states in the cavity mode is a function of position[6], leading to a variation in scattering rates shown as $\beta(\vec{r})$ in equation (11d). The orientational dependence is due to the fact that the scattering is quadratically dependent on the polarisability and it cannot be cancelled out with the denominator as the case for small particles.

To deduce the shape anisotropy in this case, it is necessary to generate a histogram of $\Phi$ for the single particle event and fit an assumed analytic probability distribution. When the nanorod is in the free diffusion with no preferred orientation, the distribution for $\theta$ is $sin\theta$ with $\theta$ in the range of $0 \sim \frac{\pi}{2}$.

$$p(\Phi)d\Phi = p(\theta)d\theta = sin\theta d\theta \quad (12)$$

With equation (11d), we could calculate $d\Phi$ and $cos\theta$

$$B(\Phi) = cos\theta = \sqrt{\frac{\Phi Re(\alpha_L-\alpha_W)-Im(\alpha_L-\alpha_W)-A\beta(\vec{r})(\alpha_L^2-\alpha_W^2)}{A\beta(\vec{r})[2Re(\alpha_L)Re(\alpha_W)+2Im(\alpha_L)Im(\alpha_W)-\alpha_L^2-\alpha_W^2]}} \quad (13)$$

Then we can get the probability distribution of the anisotropy loss tangent $\Phi$: $p(\Phi) = \frac{Re(\alpha_L-\alpha_W)}{2A\beta(\vec{r})[\alpha_L^2+\alpha_W^2-2Re(\alpha_L)Re(\alpha_W)-2Im(\alpha_L)Im(\alpha_W)]\sqrt{B(\Phi)}}$ that we used to perform the fitting in the main text. The polarisability anisotropy and related calculations are shown in the previous section (SI section 5)

Without scattering, we directly used a Gaussian instrumental response to fit the histograms of the anisotropy loss tangent $\Phi$, we could get the standard error for this fitting. Then we used the error propagation equation to calculate the standard error of $\mu$ from that of $\Phi$ with the equation $SE(\Phi) = \frac{\partial \Phi}{\partial \mu} SE(\mu)$. In the case considering the scattering, we convoluted the probability distribution of the anisotropy loss tangent $\Phi$ with the Gaussian instrumental response to perform the fitting. Thus, we employed the bootstrap method and set the percentile range of 0.68 to get the standard error of $\mu$.

### 7. Spherical particle data

To demonstrate that non-zero $S_1$ signals are introduced by the particle's anisotropy, we measure gold nanospheres of 60 nm diameter for a comparison. From the FWHM of $S_0$ signals over a 500 ms window, increases in photon loss indicating the presence of a particle are visible. Two red-rectangular areas in Figure S3(a) represent two single-particle events. The corresponding $S_0$, $S_1$ and $S_3$ traces over the same 500 ms time window are shown in in Figure S3(b), revealing that the $S_0$ and $S_3$ peaks both attenuate as a result of this increased loss while no signal is seen in the $S_1$ trace. Zooming into individual sweeps (Figure S3(c)), confirms this result.

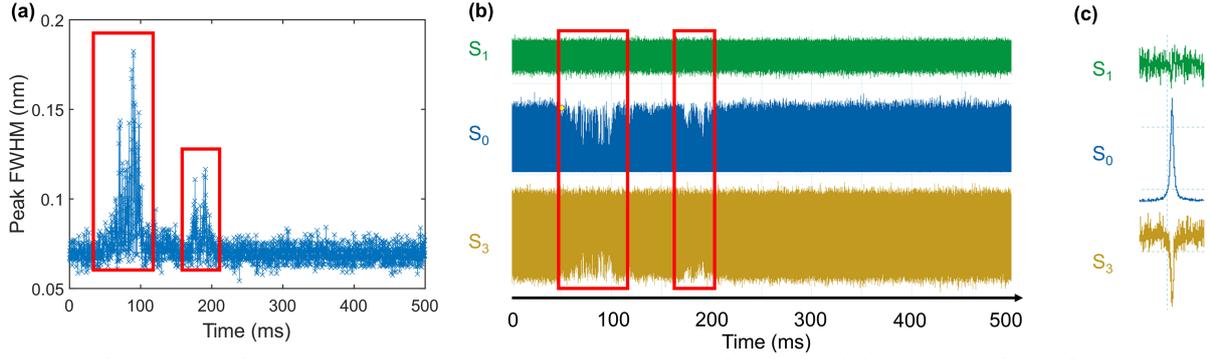

Figure S3: Data of gold nanospheres with a diameter of 60 nm. (a) FWHM of the $S_0$ peak was measured using the APD over a 500 ms window, showing two discrete single-particle events due to loss-induced broadening. (b) Raw data of $S_0$, $S_1$, and $S_3$ over this 500 ms window. (c) An example of an individual sweep after zooming in.

## 8. Rotational diffusion analysis

There are three common models to derive the rotational diffusion coefficient for rod-shaped particles, with different applicable ranges for the aspect ratio $\mu$, including Perrin theory[7], [8], [9], [10] for $\mu < 2$, Tirado and Garcia de la Torre (TG) theory[11], [12] for $2 < \mu < 30$, and Broersma's relations (BR)[12], [13] for $\mu > 5$. With our gold nanorod in the range of $\mu < 2$, we calculated with Perrin theory and also compared with TG theory (14c). With Perrin theory, equation (14a)[9] and (14b)[10] show a slightly different value:

$$D_r = \frac{3k_B T}{2\pi\eta L^3}\frac{\mu^4}{\mu^4-1}\left[\frac{(2\mu^2-1)\ln\left(\mu+\sqrt{\mu^2-1}\right)}{\mu\sqrt{\mu^2-1}}-1\right] \quad (14a)$$

$$D_r = \frac{k_B T}{4\pi\eta V}\frac{\mu^2}{\mu^4-1}\left[\frac{(2\mu^2-1)\ln\left(\mu+\sqrt{\mu^2-1}\right)}{\mu\sqrt{\mu^2-1}}-1\right] \quad (14b)$$

$$D_r = \frac{3k_B T}{\pi\eta L^3}\left(ln\mu - 0.662 + \frac{0.917}{\mu} - \frac{0.05}{\mu^2}\right) \quad (14c)$$

where $\eta$ is the viscosity of the fluid, water in this case, $V$ is the particle size, $T$ is the temperature and $k_B$ is the Boltzmann constant. With the rotational diffusion coefficient, we calculate the averaged angular displacement over 1 $\mu s$ using $<\theta^2> = 2D_r t$. The characteristic angular diffusion time $\tau = \frac{\pi^2}{8D_r}$ is also calculated. The summary is shown in Table S1.

|  | Perrin's equation ($\mu < 2$, (14a)) | Perrin's equation ($\mu < 2$, (14b)) | TG theory ($2 < \mu < 30$) |
|---|---|---|---|
| Sample A: D=32 nm, L= 48 nm | $2.74 \times 10^4$ (13.41°) | $2.35 \times 10^4$ (12.42°) | $1.45 \times 10^4$ (9.76°) |
| Characteristic time | 45 $\mu s$ | 53 $\mu s$ | 85 $\mu s$ |
| Sample B: D=28 nm, L= 40 nm | $4.43 \times 10^4$ (17.03°) | $3.85 \times 10^4$ (15.9°) | $2.35 \times 10^4$ (12.42°) |
| Characteristic time | 28 $\mu s$ | 32 $\mu s$ | 52 $\mu s$ |

Table S1. Summary of rotational diffusion with different models: rotational diffusion coefficients (the corresponding averaged angular displacement over 1 $\mu s$), and the characteristic time.

Apart from the Brownian rotational diffusion, the particle also experiences the optical torque. We calculated the maximum optical torque $M$ in a Gaussian beam under the experimental condition and calculated the corresponding angular displacement over $1 \, \mu s$ using $\theta = \frac{M*t*D_r}{k_B T}$ with the diffusion coefficient calculated from equation (14b). Angular displacements of $0.12°$ and $0.2°$ were found for samples A and B respectively, small enough to be neglected.

## 9. Thermal effects

Heating of the nanorod by the optical field could in principle result in a change of shape[14], so we calculated the temperature increase on the surface of the gold nanorod during the resonance:

$$\Delta T = \frac{C_{abs} I}{4\pi R_{eff} \kappa \beta} \quad (15a)$$

where $C_{abs}$ is the absorption cross-section, $I$ is the intensity of incident light, $R_{eff}$ is the effective radii if the particle were a sphere, $\kappa$ is the thermal conductivity of water, and $\beta = 1 + 0.96587 \ln^2(\frac{L}{2r})$ is the thermal capacitance coefficient for rod shape of the particle[15]. The absorption cross section is deduced from Gans modification to Mie theory[2], [3], [4]:

$$C_{abs} = \frac{2\pi V_p \epsilon_m^{\frac{3}{2}}}{3\lambda} \sum_{x,y,z} \frac{\left(\frac{1}{p_i^2}\right)\epsilon_2}{\{\epsilon_1 + [\frac{1-p_i}{p_i}]\epsilon_m\}^2 + \epsilon_2^2} \quad (15b)$$

where $V_p$ is the volume of the particle, $\epsilon_m$ is the permittivity of the media, $\epsilon_1$, $\epsilon_2$ are the real and imaginary part of the permittivity of gold at the wavelength $\lambda = 640 \, nm$ of the incident light, and $p_i \, (i = x, y, z)$ are the depolarisation factors along the three axes of the nanorod, which we mentioned in equation (9). From the APD voltage of the transmitted light from the empty cavity, we calculate the intra-cavity light intensity. The calculated temperature increases for sample A and B are 11K and 7K respectively. These temperature changes are too small to induce any shape change in the nanoparticles.

## 10. Relationship between $\Phi$ and $\mu$

From equation (8) and (9), we can get:

$$\alpha_L - \alpha_W = \frac{\epsilon - \epsilon_m}{\epsilon_m + p_L(\epsilon - \epsilon_m)} - \frac{\epsilon - \epsilon_m}{\epsilon_m + \frac{1-p_L}{2}(\epsilon - \epsilon_m)} \quad (16a)$$

Let $\frac{\epsilon}{\epsilon_m} = \epsilon' = C + iD$, (16a) becomes:

$$\alpha_L - \alpha_W = \frac{(C+iD)^2(1-3p_L)}{2+(1+p_L)C+p_L(1-p_L)(C^2-D^2)+iD(1+p_L)+2iCD(1-p_L)p_L} \quad (16b)$$

Therefore, the anisotropy loss tangent $\Phi$ becomes:

$$\Phi = \frac{Im(\alpha_L - \alpha_W)}{Re(\alpha_L - \alpha_W)} = \frac{D[4C + (C^2+D^2) + (C^2+D^2)p_L]}{2(C^2-D^2) + (2C+C^2+D^2-p_L^2)(C^2+D^2)} \quad (17)$$

The numerator is not a factor of the denominator, and Fig.3(e) in the main text illustrates the relationship numerically.

## 11. SEM results analysis

Assuming a random distribution of gold nanorods' orientation, the true aspect ratio can be calculated with the distribution of measured SEM results:

$$\mu = \frac{\mu_{measured} - 1}{\sin\theta} + 1 \quad (18)$$

where $\theta$ is the same as in equation (10). For each calculation, $\mu_{measured}$ is randomly selected from the Gaussian distribution of measured SEM results. The sample size was set to 100,000 to improve the accuracy of the final result shown in Figure S4.

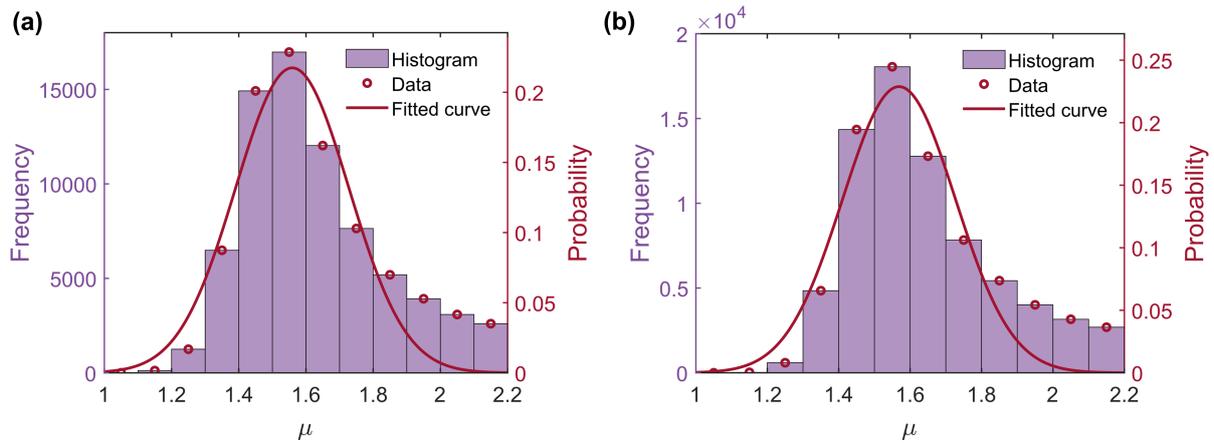

Figure S4: SEM simulation results assuming random orientations on the image plane. (a) and (b) are results for samples A and B respectively.

## 12. Minimum aspect ratio analysis

The noise level of the balanced photodiode is around $\pm 50\ mV$. With the curve fitting of our experimental data, we could get an averaged value of 0.78 for $4F^2$ in equation (6c). We then look at the case when $\psi = 0°$ and $\Delta = 0$ and we ignore the scattering impact, $\gamma = \frac{w_c \alpha_i I(\vec{r})}{2V_m}$, $\varepsilon = \frac{w_c \Delta\alpha_r I(\vec{r})}{4V_m}$, $\eta = \frac{w_c \Delta\alpha_i I(\vec{r})}{4V_m}$ where $\alpha_i$ is the imaginary part of the polarisability if the particle were a sphere and all the other parameters are the same as in equation (11a). With all these inserted to equation (6c), we could find that the magnitude of $S_1$ is impacted by $V_m$ and $I(\vec{r})$. Then we insert the rough estimation of these two from the averaged value from the fitting of our experimental data, then the equation (6c) will only have unknown values for $\Delta\alpha_r$ and $\Delta\alpha_i$, and with equation (8) and (9) mentioned above, we could get the value of 1.05 for the minimum aspect ratio we could measure with our apparatus in this case.

## References

[1] B. Schaefer, E. Collett, R. Smyth, D. Barrett, and B. Fraher, "Measuring the Stokes polarization parameters," *Am J Phys*, vol. 75, no. 2, pp. 163–168, 2007, https://doi.org/10.1119/1.2386162.

[2] R. Gans, "Über die Form ultramikroskopischer Silberteilchen," *Ann Phys*, vol. 352, no. 10, pp. 270–284, 1915, https://doi.org/10.1002/andp.19153521006.

[3] R. Gans, "Über die Form ultramikroskopischer Goldteilchen," *Ann Phys*, vol. 342, no. 5, pp. 881–900, 1912, https://doi.org/10.1002/andp.19123420503.

[4] G. C. Papavassiliou, "Optical properties of small inorganic and organic metal particles," *Progress in Solid State Chemistry*, vol. 12, no. 3–4, pp. 185–271, 1979, https://doi.org/10.1016/0079-6786(79)90001-3.

[5]   P. B. Johnson and R. W. Christy, "Optical Constants of the Noble Metals," *Phys Rev B*, vol. 6, no. 12, pp. 4370–4379, 1972, https://doi.org/10.1103/PhysRevB.6.4370.

[6]   A. A. P. Trichet, P. R. Dolan, D. James, G. M. Hughes, C. Vallance, and J. M. Smith, "Nanoparticle Trapping and Characterization Using Open Microcavities," *Nano Lett*, vol. 16, no. 10, pp. 6172–6177, 2016, https://doi.org/10.1021/acs.nanolett.6b02433.

[7]   F. Perrin, "Mouvement brownien d'un ellipsoide - I. Dispersion diélectrique pour des molécules ellipsoidales," *Journal de Physique et le Radium*, vol. 5, no. 10, pp. 497–511, 1934, https://doi.org/10.1051/jphysrad:01934005010049700.

[8]   S. H. Koenig, "Brownian motion of an ellipsoid. A correction to Perrin's results," *Biopolymers*, vol. 14, no. 11, pp. 2421–2423, 1975, https://doi.org/10.1002/bip.1975.360141115.

[9]   A. H. Kumar, S. J. Thomson, T. R. Powers, and D. M. Harris, "Taylor dispersion of elongated rods," *Phys Rev Fluids*, vol. 6, no. 9, p. 094501, 2021, https://doi.org/10.1103/PhysRevFluids.6.094501.

[10]  Y. Han, A. Alsayed, M. Nobili, and A. G. Yodh, "Quasi-two-dimensional diffusion of single ellipsoids: Aspect ratio and confinement effects," *Phys Rev E*, vol. 80, no. 1, p. 011403, 2009, https://doi.org/10.1103/PhysRevE.80.011403.

[11]  M. M. Tirado and J. G. de la Torre, "Rotational dynamics of rigid, symmetric top macromolecules. Application to circular cylinders," *J Chem Phys*, vol. 73, no. 4, pp. 1986–1993, 1980, https://doi.org/10.1063/1.440288.

[12]  R. Nixon-Luke and G. Bryant, "A Depolarized Dynamic Light Scattering Method to Calculate Translational and Rotational Diffusion Coefficients of Nanorods," *Particle & Particle Systems Characterization*, vol. 36, no. 2, p. 1800388, 2019, https://doi.org/10.1002/ppsc.201800388.

[13]  S. Broersma, "Viscous Force Constant for a Closed Cylinder," *J Chem Phys*, vol. 32, no. 6, pp. 1632–1635, 1960, https://doi.org/10.1063/1.1730995.

[14]  S.-S. Chang, C.-W. Shih, C.-D. Chen, W.-C. Lai, and C. R. Chris Wang, "The Shape Transition of Gold Nanorods," *Langmuir*, vol. 15, no. 3, pp. 701–709, 1999, https://doi.org/10.1021/la980929l.

[15]  G. Baffou, R. Quidant, and F. J. García de Abajo, "Nanoscale Control of Optical Heating in Complex Plasmonic Systems," *ACS Nano*, vol. 4, no. 2, pp. 709–716, 2010, https://doi.org/10.1021/nn901144d.